\documentclass[twocolumn,aps,prl,floatfix,superscriptaddress]{revtex4-2}
\usepackage{amsmath,amssymb}
\usepackage{graphicx,float}
\usepackage{times,txfonts}
\usepackage{color}
\usepackage{multirow}
\usepackage{bbold}
\usepackage{color}
\usepackage{soul}
\usepackage{hyperref}

\hypersetup{
   pdfpagemode=None,
   pdfstartpage=1,
   pdfmenubar=true,
   pdftoolbar=true,
   colorlinks = true,
   linkcolor=blue,
   citecolor=blue,
   urlcolor=blue,
   bookmarksopen=false
 }

\usepackage{times,txfonts}
\usepackage{natbib}
\usepackage{nicefrac}
\usepackage{blindtext}
\usepackage[export]{adjustbox}
\usepackage{mathrsfs}
\usepackage{textcomp}
\usepackage{blkarray}
\usepackage{multirow}
\usepackage[normalem]{ulem}
\usepackage[T1]{fontenc}
\usepackage[utf8]{inputenc}
\usepackage{color}
\usepackage{babel}
\usepackage{amsmath}
\usepackage{graphicx}
\usepackage{esint}

\newcommand{\abs}[1]{\left| #1 \right|}

\newcommand{\bra}[1]{\left\langle #1 \right|}
\newcommand{\ket}[1]{\left| #1 \right\rangle}

\renewcommand{\epsilon}{\varepsilon}

\def\VR{\kern-\arraycolsep\strut\vrule &\kern-\arraycolsep}
\def\vr{\kern-\arraycolsep & \kern-\arraycolsep}
\usepackage{sistyle}
\usepackage{physics}
\begin{document}
\title{Manifestation of the Berry connection in chiral lattice systems}

\author{Francesco Di Colandrea}
\affiliation{Nexus for Quantum Technologies, University of Ottawa, K1N 5N6, Ottawa, ON, Canada}

\author{Nazanin Dehghan}
\affiliation{Nexus for Quantum Technologies, University of Ottawa, K1N 5N6, Ottawa, ON, Canada}
\affiliation{National Research Council of Canada, 100 Sussex Drive, Ottawa ON Canada, K1A 0R6}
    
\author{Filippo Cardano}
\affiliation{Dipartimento di Fisica ``Ettore Pancini,'' Università degli Studi di Napoli Federico II, Complesso Universitario di Monte Sant'Angelo, via Cintia, 80126 Napoli, Italy}

\author{Alessio D'Errico}
\email[Correspondence email address: ]{aderrico@uottawa.ca}
\affiliation{Nexus for Quantum Technologies, University of Ottawa, K1N 5N6, Ottawa, ON, Canada}
\affiliation{National Research Council of Canada, 100 Sussex Drive, Ottawa ON Canada, K1A 0R6}
    
\author{Ebrahim Karimi}
\affiliation{Nexus for Quantum Technologies, University of Ottawa, K1N 5N6, Ottawa, ON, Canada}
\affiliation{National Research Council of Canada, 100 Sussex Drive, Ottawa ON Canada, K1A 0R6}
    

\begin{abstract}
The Aharonov-Bohm effect is a physical phenomenon where the vector potential induces a phase shift of electron wavepackets in regions with zero magnetic fields. It is often referred to as evidence for the physical reality of the vector potential. A similar effect can be observed in solid-state systems, where the Berry connection can influence electron dynamics. Here, we show that in chiral-symmetric processes the Berry connection 
determines an observable effect on the mean chiral displacement of delocalized wavefunctions. This finding is supported by a photonic experiment realizing a topological quantum walk, and demonstrates a new effect that can be attributed directly to the presence of a gauge field.  
\end{abstract}


\maketitle

\noindent\textit{Introduction:} Gauge fields play a fundamental role in modern physics. Besides being key mathematical objects for electromagnetism, quantum field theories, and electronic band structure, there has been a long debate on whether gauge fields should constitute the fundamental elements of the theory. In 1959, Y. Aharonov and D. Bohm~\cite{aharonov1959significance} showed that electron wavefunctions can experience phase shifts induced by the vector potential even when crossing regions of space with zero electric or magnetic fields. This effect has been later verified in several experiments~\cite{tonomura1982observation,webb1985observation,matteucci1985new, tonomura1986evidence, timp1987observation, allman1992scalar, bachtold1999aharonov, haug2019aharonov}, and can be understood as a geometric-phase effect due to the presence of an obstruction in space~\cite{yau2002aharonov,xiao2010berry,cohen2019geometric}. Similar phenomena have been observed for water waves around flux vortices~\cite{berry1980wavefront} and in optical systems~\cite{li2014photonic, parto2019observation}. The Aharonov-Bohm (AB) effect thus stimulated an ongoing debate on the physical nature of electromagnetic potentials~\cite{vaidman2012role,aharonov2015comment,vaidman2015reply,aharonov2016nonlocality,li2022gauge, paiva2023coherence}. 

The modern electronic band theory in solid-state physics predicts a plethora of phenomena associated with quantities analogous to the magnetic field and the vector potential in electromagnetism. These quantities are the Berry curvature~$\boldsymbol{\Omega}$ and the Berry connection $\boldsymbol{\mathcal{A}}$~\cite{xiao2010berry}, respectively. The Berry curvature is a gauge-invariant field arising from the geometry of band eigenstates in the quasi-momentum space, which can introduce anomalous velocity effects to the motion of wavepackets when an external field is applied on the system~\cite{changniu1995, chang1996berry,xiao2010berry,price2016measurement, wimmer2017experimental, d2020two}. The integral of the Berry curvature over the quasi-momentum space gives the topological invariant of Chern insulators~\cite{bernevig2013topological,asboth2016berry,d2020two}. 
This analogy can be exploited to design AB experiments in lattice systems, with the role of the vector potential played by the Berry connection. For instance, the AB effect has been observed in an atomic simulator of graphene, where $\boldsymbol{\Omega}$ is zero everywhere and singular in correspondence of the Dirac points~\cite{duca2015aharonov}. 
\\
In this work, we report on a novel effect of the Berry connection on the motion of wavepackets. We focus on tight-binding models exhibiting chiral symmetry, and analyze the temporal evolution of wavepackets sharply peaked in the quasi-momentum space. Chiral symmetry is characterized by the existence of a unitary operator that pairs states with opposite energy (measured with respect to the Fermi level), and is typical of systems with two -- or an even number of -- sites per unit cell, such as models of polyacetylene chains~\cite{su1980soliton, asboth2016berry} and graphene~\cite{RevModPhys.81.109}. We evaluate the time evolution of the mean chiral displacement (MCD)~\cite{cardano2017detection, maffei2018topological, d2020bulk}, which gives the weighted difference in the mean position of the wavepacket distribution on the two sublattices.
Within a specific choice of the gauge, having a distinct geometric interpretation, an effect of the Berry connection $\boldsymbol{\mathcal{A}}$ on the MCD of wavepackets is demonstrated: in the long-time limit, the MCD converges to the convolution between the Berry connection and the initial wavepacket distribution in the reciprocal lattice, that is the quasi-momentum space.

As detailed below, the gauge choice guarantees a direct geometric interpretation of the Berry connection in terms of the eigenstates, when represented as unit vectors on the Bloch sphere. 
This result is firstly verified with numerical simulations of different prototypical solid-state models, then experimentally observed in a photonic quantum walk where the lattice is encoded in the light transverse wavevector and the internal degree of freedom (sublattice) is associated with the optical polarization \cite{d2020two}. A chiral-symmetric unitary evolution operator is implemented via patterned anisotropic devices.  
It is shown that the time-averaged MCD can be used to measure the Berry connection in one-dimensional (1D) topological quantum walks and extract the corresponding topological invariant.\\
\begin{figure}[!h]
    \centering
    \includegraphics[width=0.85\columnwidth]{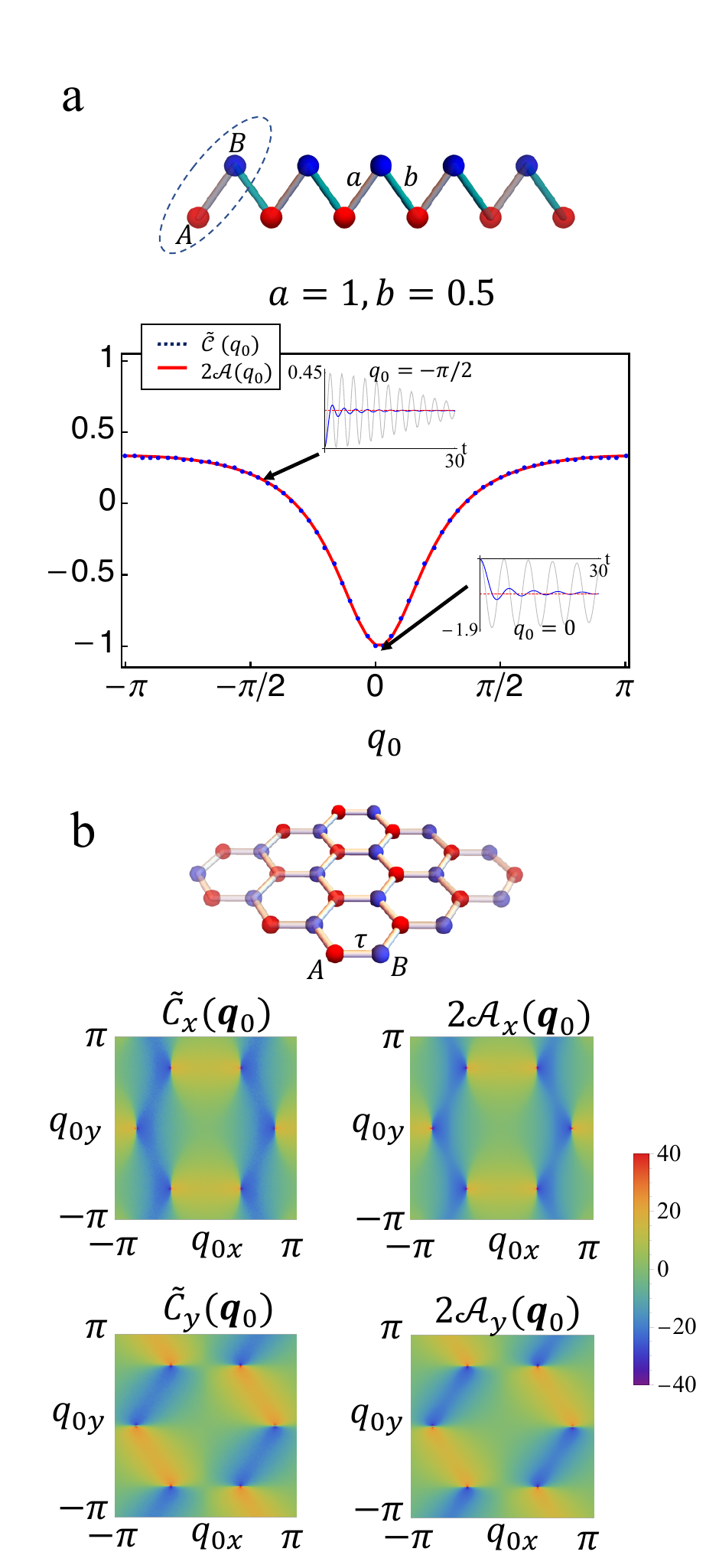}
  \caption{\textbf{Mean Chiral Displacement and Berry connection in chiral models.} a.~Time-averaged MCD for the SSH model (top), evaluated for wavepackets (${w=0.1}$) centered in different points of the Brillouin zone (blue points), and compared with the Berry connection (red curve) after $t=30/a$. Insets show the dynamical evolution of the MCD (gray lines), and its average in time (blue line), for selected quasi-momentum values. b.~Comparison of time-averaged MCD ($50/\tau\leq t\leq 60/\tau$, where $\tau$ is the hopping amplitude between the two sublattices, ${w=\pi/10}$) and Berry connection in two-band graphene (top). Note that the color scale is truncated to the maximum of the MCD since $\mathbf{\mathcal{A}}$ diverges in the proximity of the Dirac cones.}
    \label{fig:MCDth}
\end{figure}

\noindent\textit{Theory:} Consider a lattice system of arbitrary spatial dimension $D$ formed by two sublattices, that is with two sites per unit cell. Precisely, the quantum states are elements of the Hilbert space $\mathcal{H}_{\ell}\otimes \mathcal{H}_s$, where $\mathcal{H}_\ell$ is spanned by eigenstates of the lattice position and dim$(\mathcal{H}_s)=2$. 

Let us assume that the system possesses chiral symmetry, i.e., there exists a unitary operator $\hat{\Gamma}$, acting on $\mathcal{H}_s$, such that $\Gamma\hat{U}=\hat{U}^{-1}\Gamma$, where $\hat{U}$ is the unitary describing the single-particle evolution. Without loss of generality, one can choose $\hat{\Gamma}=\hat{\sigma}_z:=\text{diag}[1,-1]$. Translation invariance and chiral symmetry allow one to write $\hat{U}$ in the block-diagonal form
\begin{equation}
    \hat{U}=\int_{\text{BZ}}\frac{d^D q}{(2\pi)^D}\,\,\mathcal{U}(\mathbf{q})\otimes\ketbra{\mathbf{q}} \label{chiralU},
\end{equation}
where
\begin{align}
    \mathcal{U}(\mathbf{q})&=\exp(-i E(\mathbf{q})\,\mathbf{n}(\mathbf{q})\cdot \hat{\boldsymbol{\sigma}}),\label{evU}\\
    \mathbf{n}(\mathbf{q})&=(n_x(\mathbf{q}),n_y(\mathbf{q}),0).
    \label{n}
\end{align}
In Eqs.~\eqref{chiralU}-\eqref{evU}, $\mathbf{q}$ is the quasi-momentum, defined in the adimensional Brillouin Zone $(\text{BZ})$ --- for square lattices, ${\text{BZ}=[-\pi,\pi[^{\otimes D}}$ ---, $\mathcal{U}$ is an operator acting on $\mathcal{H}_s$, with eigenvalue $E$ and eigenstate $\mathbf{n}$, with ${\abs{\mathbf{n}}=1}$, and ${\hat{\boldsymbol{\sigma}}=(\sigma_x,\sigma_y, \sigma_z)}$ is the vector of the three Pauli matrices. Equation~\eqref{n} is a consequence of chiral symmetry: the eigenstates are represented on the Bloch sphere as unit vectors $\mathbf{n}(\mathbf{q})$,  which lie in a plane perpendicular to the axis specified by the chiral operator (here chosen as the $z$ axis). 

In 1D lattices, chiral symmetry allows defining a topological invariant, the winding number $\nu$, associated with the number of times the vector $\mathbf{n}(\mathbf{q})$  spans the plane when the quasi-momentum runs across one $\text{BZ}$. It has been proven that the MCD provides an observable quantity tracking the topological invariant of 1D chiral systems  ~\cite{cardano2017detection,maffei2018topological, meier2018observation, haller2020detecting, st2021measuring, d2020bulk}.

For two-band models, the MCD is defined as
\begin{align}
    \mathcal{C}_i(t)=2\bra{\psi (t)}\hat{\Gamma}\hat{x}_i\ket{\psi(t)},
\label{eq:MCD}
\end{align}
where $\hat{x}_i$ is the $i$-th component of the lattice position operator and $\ket{\psi(t)}=\hat{U}^t\ket{\psi(0)}$, $\ket{\psi(0)}$ being the system wavefunction at time ${t=0}$. Equation~\eqref{eq:MCD} can be interpreted as the weighted difference between the mean positions on the two sublattices (where the weights are given by the probability of being in either sublattice). The MCD asymptotically converges to the winding number if $\ket{\psi(0)}$ is a localized state \cite{cardano2017detection} or, more generally, can be mapped to a localized state via a translation-invariant unitary operator~\cite{d2020bulk}. 

In this work, a different scenario is considered, where the initial state is a wavepacket exhibiting a narrow distribution in the reciprocal lattice, ${\ket{\psi(0)}=\int_\text{{BZ}}d^Dq/(2\pi)^D\, G_{w,\mathbf{q}_0}(\mathbf{q})\ket{\mathbf{q}}\otimes\ket{\phi_0}}$, where $\ket{\phi_0}$ is a sublattice state and $G_{w,\mathbf{q}_0}(\mathbf{q})$ a function peaked around $\mathbf{q}_0$ with characteristic width $w$. We obtain -- see Supplementary Material for the detailed derivation--
%

\begin{equation}
\begin{split}
    \mathcal{C}_i(t)&=2\int_\text{BZ}\frac{d^D q}{(2\pi)^D}\abs{G_{w,\mathbf{q}_0}}^2\sin^2(tE)2\mathcal{A}_i\\ 
    &=\int_\text{BZ}\frac{d^D q}{(2\pi)^D}\abs{G_{w,\mathbf{q}_0}}^22\mathcal{A}_i\left(1-\cos(2Et)\right),
\end{split}
\label{eq:MCDtot}
\end{equation}
where the index $i$ refers to different spatial components, $\mathbf{v}_{\Gamma}$ is the \textit{chiral vector}, i.e., ${\mathbf{v}_{\Gamma}\cdot\mathbf{n}(\mathbf{q})=0}$ for each $\mathbf{q}$, and ${\mathcal{A}_i(\mathbf{q})=\left(\mathbf{n}(\mathbf{q})\cross \partial_{q_i}\mathbf{n}(\mathbf{q})\right)\cdot \mathbf{v}_{\Gamma}/2}$. This quantity, being proportional to the solid angle enclosed by the vectors $\mathbf{v}_{\Gamma},\, \mathbf{n}(\mathbf{q})$, and ${\mathbf{n}(\mathbf{q}+d\mathbf{q})}$ in the unit sphere \cite{bliokh2019geometric}, captures a geometric property of the system. $\mathcal{A}_i(\mathbf{q})$ is equal to the Berry connection ${\mathcal{A}_i(\mathbf{q}):=\bra{\mathbf{n}(\mathbf{q})}i\nabla_{\mathbf{q}}\ket{\mathbf{n}(\mathbf{q})}}$ in the gauge where the system eigenstates are written as ${\ket{\mathbf{n}(\mathbf{q})}=(e^{-i\phi(\mathbf{q})}\ket{\uparrow}+\ket{\downarrow})/\sqrt{2}}$, with $\phi:=\arctan{(n_y/n_x)}$ and $\hat{\Gamma}=:\ketbra{\uparrow}-\ketbra{\downarrow}$ -- recall that a gauge transformation can be defined by a local phase transformation on the quantum states ${\ket{\psi}\rightarrow e^{i f(\mathbf{q})}\ket{\psi}}$.
The last integral in Eq.~\eqref{eq:MCDtot} gives an oscillating contribution that generally decreases in amplitude as $\sim 1/\sqrt{t}$, and thus, asymptotically, we obtain
\begin{align}
    \mathcal{C}_i&\sim\int_\text{BZ}\frac{d^D q}{(2\pi)^D}\, \abs{G_{w,\mathbf{q}_0}(\mathbf{q})}^2 2\mathcal{A}_i(\mathbf{q}).
\label{eq:Ci}
\end{align}
%


For initial states narrowly peaked in the reciprocal lattice ($w\rightarrow 0$), the MCD can thus probe the local value of the Berry connection:
\begin{align}
\mathcal{C}_i\sim 2\mathcal{A}_i(\mathbf{q}).
\end{align}
 

If ${\abs{G_{w,\mathbf{q}_0}}^2=g_w(\mathbf{q}-\mathbf{q}_0)}$, Equation~\eqref{eq:MCDtot} can be seen as the convolution between the function $g_w$ and $\sin^2(tE)\mathcal{A}_i$. Hence, one can extract $\sin^2(tE)\mathcal{A}_i$ for finite widths $w$ via a deconvolution analysis of the measured $\mathcal{C}_i$ as a function of $\mathbf{q}_0$. The factor $\sin^2(tE)$ is either a constant in the case of flat bands or averages to $1/2$ in the long-time limit. The Berry connection can thus be extracted also in situations with evolutions limited in time by extracting the time average of the deconvolved MCD. 

A comparison between the time-averaged MCD $\tilde{\mathcal{C}}$ of two typical chiral models is shown in Fig.~\ref{fig:MCDth}. In both cases, the initial state is a Gaussian wavepacket sharply peaked around a given quasi-momentum value $\mathbf{q}_0$, ${G_{w,\mathbf{q}_0}=\mathcal{N}\exp(-(\mathbf{q}-\mathbf{q}_0)^2/w^2)}$, where $\mathcal{N}$ is a normalization factor. In panel a, the Su-Schrieffer-Heeger (SSH) model for a composite 1D lattice is considered~\cite{su1980soliton}. The parameters $a$ and $b$ correspond to the intracell and intercell hopping amplitudes, respectively. The MCD is evaluated for different values of $q_0$ spanning the BZ. The agreement between $\tilde{\mathcal{C}}$ and the Berry connection is evident after $t=30/a$. In Fig.~\ref{fig:MCDth}b, the same simulation is performed for the two-band tight-binding graphene. The time-averaged MCD converges to $2\boldsymbol{\mathcal{A}}$. Although is not evident in the figure, this convergence fails in the extreme vicinity of the Dirac points, where $2\boldsymbol{\mathcal{A}}$ diverges and the MCD falls to zero, as shown in the Supplementary Material.\\
\begin{figure*}
    \centering
    \includegraphics[width=\textwidth]{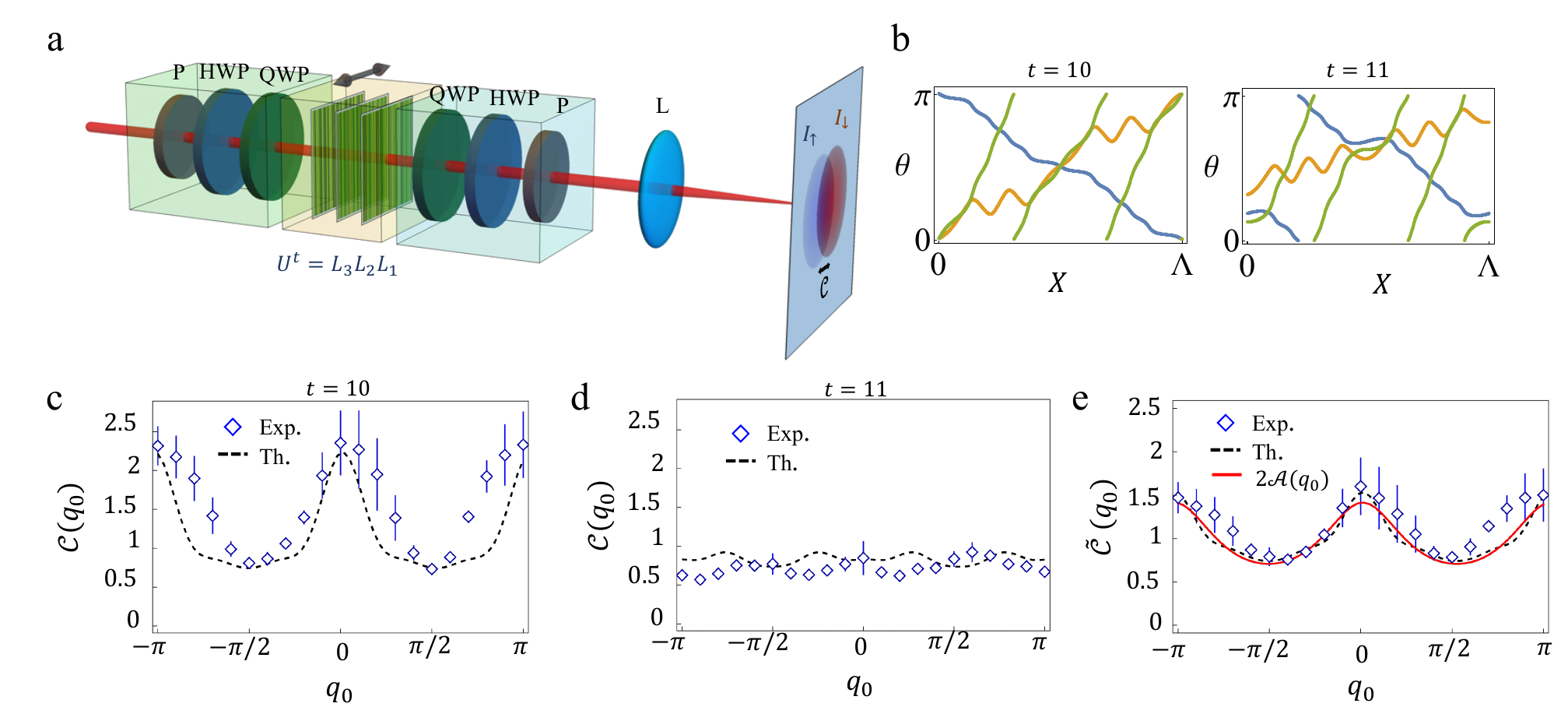}
  \caption{\textbf{MCD induced by the Berry connection in a photonic quantum walk.} a. Sketch of the experimental setup. The laser beam is prepared in an arbitrary polarization state, set by a polarizer (P), a half-wave plate (HWP) and a quarter-wave plate (QWP). To simulate wavepacket dynamics, the beam waist is adjusted so as $w_0<\Lambda$. The overall chiral-symmetric evolution operator, $U^t$, is implemented via three liquid-crystal metasurfaces. A rigid translation of the bulk of plates along the $X$ direction is equivalent to changing the value $q_0$ for the wavepacket. After the QW, a projection onto the chiral eigenstates, $\ket{\uparrow}$ and $\ket{\downarrow}$, is performed, and the resulting intensity is recorded in the focal plane of a lens (L), corresponding to the lattice space of the QW. The difference in the centroid position of the two intensity distributions, $I_{\uparrow}$ and $I_{\downarrow}$, gives the MCD. b. Optic-axis patterns $\theta(X)$ of the liquid-crystal plates $L_{1,2,3}$ implementing $t=10$ and $t=11$ steps. The distance ${\Lambda=0.25}$ cm corresponds to the largest spatial period and defines a BZ. The lattice spacing in the transverse-wavevector domain is thus $2\pi/\Lambda$. c-d.~The measured MCD at ${t=10}$ and ${t=11}$ is compared with theoretical predictions. e.~The time-averaged MCD obtained experimentally is compared with the average computed from a complete simulation of the ideal QW evolution (dashed curve) and with the Berry connection $\mathcal{A}(q)$ (red curve).}
    \label{fig:MCDexp}
\end{figure*}

\noindent\textit{Experimental Results:} The effect of the Berry connection on the MCD of wavepackets is experimentally investigated in a photonic quantum walk (QW). In this experiment, the sublattice degree of freedom is encoded into the polarization of light. The lattice can be encoded in an infinite-dimensional degree of freedom, such as the (discretized) transverse wavevector \cite{d2020two,d2021bloch,esposito2022quantum}. Following the approach recently devised in Ref.~\cite{di2023ultra}, the chiral-symmetric unitary is implemented by a minimal stack of three patterned waveplates, specifically a half-wave plate sandwiched between two quarter-wave plates. This scheme allows implementing a space-dependent polarization transformation which maps into the target evolution~\cite{di2023ultra}. 

The chiral process considered here is the QW introduced in Ref.~\cite{cardano2017detection}. The unit step $U=T\cdot W$ consists of a polarization rotation $W:=(\sigma_0+i\sigma_x)/\sqrt{2}$, followed by a polarization-dependent lattice translation $T:=\sum_x (\ket{x+1}\bra{x}\otimes\sigma_- + \ket{x-1}\bra{x}\otimes\sigma_+)$, where $\sigma_{\pm}=\sigma_x\mp i\sigma_y$. Here it is assumed that the positive and negative eigenstates of $\sigma_z$ are left and right circular polarizations, $\ket{L}$ and $\ket{R}$, respectively. Numerical simulations show that it is sufficient to consider 10-step and 11-step processes, i.e., $U^{10}$ and $U^{11}$, to observe a convergence of the time-averaged MCD to the local Berry connection. While long evolutions are typically achieved by stacking single-step waveplates~\cite{cardano2017detection, d2020bulk, d2021bloch}, here the more compact approach of designing only three plates implementing either ${t=10}$ or ${t=11}$ steps is adopted.  

A simplified experimental setup is shown in Fig.~\ref{fig:MCDexp}a. A collimated laser beam (${\lambda=810}$ nm) is prepared in an arbitrary polarization state and propagates through three patterned waveplates implementing $U^t$. A projection is then performed onto the two eigenstates of the chiral operator:
\begin{eqnarray}
\ket{\uparrow}&=&\cos\left(\frac{\pi}{8}\right)\ket{L}-i\sin\left(\frac{\pi}{8}\right)\ket{R},\\
\ket{\downarrow}&=&\sin\left(\frac{\pi}{8}\right)\ket{L}+i\cos\left(\frac{\pi}{8}\right)\ket{R}.
\end{eqnarray}
The lattice space is accessed via an optical Fourier transform, i.e., by measuring the intensity distribution $I_{\uparrow/\downarrow}(x)$ in the focal plane of a lens placed after the projection stage. The MCD is experimentally retrieved by measuring the weighted difference between the center of mass of the intensity distributions of the two chiral projections: ${\mathcal{C}=2\sum_x\,(I_{\uparrow}(x)-I_{\downarrow}(x))x }$, where it is assumed that the intensities are normalized according to ${\sum_x(I_{\uparrow}(x)+I_{\downarrow}(x))=1}$. Here, with a slight abuse of notation, we identify the coordinate $x$ at the lens focus with the index specifying the correct mapping. This identification is obtained by expressing $x$ in units of the lattice spacing. The latter depends on the inverse of the characteristic period ${\Lambda=0.25}$ cm of the plates implementing the QW evolution and the focal length ${f=25~\text{cm}}$ of the lens implementing the Fourier transform. The distance $\Lambda$ also corresponds to the physical extension of a single BZ in our setup~\cite{d2020two}. In this implementation, the lattice sites are mapped into states carrying $x$ units of transverse momentum $2\pi/\Lambda$, and as a consequence, the quasi-momentum $q$ corresponds to the transverse position $X$ -- see the Supplementary Material and Ref.~\cite{d2020two} for more details. The plates implementing the QW are liquid-crystal metasurfaces exhibiting a position-dependent optic-axis orientation $\theta(X)$, plotted in Fig.~\ref{fig:MCDexp}b. The optical retardation $\delta$ of these devices is uniform but can be tuned. The first and last plates, $L_1$ and $L_3$, act as quarter-wave plates, ${\delta_{1,3}=\pi/2}$, while the intermediate plate, $L_2$, acts as a half-wave plate, ${\delta_2=\pi}$.
In this setup, the width $w$ of the initial wavefunction in the reciprocal lattice is proportional to the laser beam waist $w_0$. With a large waist, ${w_0\geq\Lambda}$, we measure a \textit{global} MCD, which asymptotically yields the topological invariant ${\nu=1}$ associated with the chosen QW~\cite{cardano2017detection}. For ${w_0<\Lambda}$, we can locally sample the BZ. We used ${w_0\sim 0.13\, \Lambda}$ (see Supplementary Material for further details on the experimental parameters). The value of $q_0$ is simply changed by translating the three metasurfaces in the $X$ direction, thus introducing an effective transverse displacement with respect to the beam propagation. The reciprocal lattice is sampled in steps of ${\Delta X=0.12}$~mm, for a total of ${n=\Lambda/\Delta X=21}$ points.
Figures \ref{fig:MCDexp}c-d show the measured MCD for ${t=10}$ and ${t=11}$ steps, respectively, compared with theoretical curves evaluated from Eq.~\eqref{eq:MCDtot}. Errors are reported as the mean standard error of four repeated measurements. In Fig.~\ref{fig:MCDexp}e, the Berry connection is compared with the average ${\tilde{\mathcal{C}}(q_0):=(\mathcal{C}(q_0,t=10)+\mathcal{C}(q_0,t=11))/2}$. We observe a good agreement with the theory. Some deviations can be ascribed to imperfections in the fabrication process and relative misalignment of the plates. Note that the measured $\tilde{\mathcal{C}}(q_0)$, when integrated over the BZ, yields ${\nu=1.15\pm 0.19}$, fully compatible with the expected value of the topological invariant.

\noindent\textit{Discussion and conclusions:} It has been shown that the Berry connection affects the relative spatial distribution of wavepackets on the two sublattices of chiral-symmetric systems. This effect is captured by the MCD. This result has been verified numerically on solid-state models, specifically the SSH model and graphene, and experimentally in a 1D photonic quantum walk.
In analogy with the AB effect, the MCD can be non-zero in regions where the Berry curvature vanishes. In the particular case of 1D systems, where the Berry curvature is not defined, this effect is still observed and proportional to the Berry connection. This is an example of a physical effect of a gauge field in a theory where gauge-invariant fields are not present. These findings can help address the fundamental question of the physical nature of gauge potentials. Moreover, our work offers a new method to measure the Berry connection of unitary processes on lattice systems, from which geometrical and topological features can be extracted.

\noindent\textit{Acknowledgements:} This work was supported by the Canada Research Chair (CRC) Program, NRC-uOttawa Joint Centre for Extreme Quantum Photonics (JCEP) via the Quantum Sensors Challenge Program at the National Research Council of Canada, and Quantum Enhanced Sensing and Imaging (QuEnSI) Alliance Consortia Quantum grant.

\bibliography{bibliography.bib}

\begin{thebibliography}{41}%
\makeatletter
\providecommand \@ifxundefined [1]{%
 \@ifx{#1\undefined}
}%
\providecommand \@ifnum [1]{%
 \ifnum #1\expandafter \@firstoftwo
 \else \expandafter \@secondoftwo
 \fi
}%
\providecommand \@ifx [1]{%
 \ifx #1\expandafter \@firstoftwo
 \else \expandafter \@secondoftwo
 \fi
}%
\providecommand \natexlab [1]{#1}%
\providecommand \enquote  [1]{``#1''}%
\providecommand \bibnamefont  [1]{#1}%
\providecommand \bibfnamefont [1]{#1}%
\providecommand \citenamefont [1]{#1}%
\providecommand \href@noop [0]{\@secondoftwo}%
\providecommand \href [0]{\begingroup \@sanitize@url \@href}%
\providecommand \@href[1]{\@@startlink{#1}\@@href}%
\providecommand \@@href[1]{\endgroup#1\@@endlink}%
\providecommand \@sanitize@url [0]{\catcode `\\12\catcode `\$12\catcode `\&12\catcode `\#12\catcode `\^12\catcode `\_12\catcode `\%12\relax}%
\providecommand \@@startlink[1]{}%
\providecommand \@@endlink[0]{}%
\providecommand \url  [0]{\begingroup\@sanitize@url \@url }%
\providecommand \@url [1]{\endgroup\@href {#1}{\urlprefix }}%
\providecommand \urlprefix  [0]{URL }%
\providecommand \Eprint [0]{\href }%
\providecommand \doibase [0]{https://doi.org/}%
\providecommand \selectlanguage [0]{\@gobble}%
\providecommand \bibinfo  [0]{\@secondoftwo}%
\providecommand \bibfield  [0]{\@secondoftwo}%
\providecommand \translation [1]{[#1]}%
\providecommand \BibitemOpen [0]{}%
\providecommand \bibitemStop [0]{}%
\providecommand \bibitemNoStop [0]{.\EOS\space}%
\providecommand \EOS [0]{\spacefactor3000\relax}%
\providecommand \BibitemShut  [1]{\csname bibitem#1\endcsname}%
\let\auto@bib@innerbib\@empty
\bibitem [{\citenamefont {Aharonov}\ and\ \citenamefont {Bohm}(1959)}]{aharonov1959significance}%
  \BibitemOpen
  \bibfield  {author} {\bibinfo {author} {\bibfnamefont {Y.}~\bibnamefont {Aharonov}}\ and\ \bibinfo {author} {\bibfnamefont {D.}~\bibnamefont {Bohm}},\ }\href {https://doi.org/10.1103/PhysRev.115.485} {\bibfield  {journal} {\bibinfo  {journal} {Phys. Rev.}\ }\textbf {\bibinfo {volume} {115}},\ \bibinfo {pages} {485} (\bibinfo {year} {1959})}\BibitemShut {NoStop}%
\bibitem [{\citenamefont {Tonomura}\ \emph {et~al.}(1982)\citenamefont {Tonomura}, \citenamefont {Matsuda}, \citenamefont {Suzuki}, \citenamefont {Fukuhara}, \citenamefont {Osakabe}, \citenamefont {Umezaki}, \citenamefont {Endo}, \citenamefont {Shinagawa}, \citenamefont {Sugita},\ and\ \citenamefont {Fujiwara}}]{tonomura1982observation}%
  \BibitemOpen
  \bibfield  {author} {\bibinfo {author} {\bibfnamefont {A.}~\bibnamefont {Tonomura}}, \bibinfo {author} {\bibfnamefont {T.}~\bibnamefont {Matsuda}}, \bibinfo {author} {\bibfnamefont {R.}~\bibnamefont {Suzuki}}, \bibinfo {author} {\bibfnamefont {A.}~\bibnamefont {Fukuhara}}, \bibinfo {author} {\bibfnamefont {N.}~\bibnamefont {Osakabe}}, \bibinfo {author} {\bibfnamefont {H.}~\bibnamefont {Umezaki}}, \bibinfo {author} {\bibfnamefont {J.}~\bibnamefont {Endo}}, \bibinfo {author} {\bibfnamefont {K.}~\bibnamefont {Shinagawa}}, \bibinfo {author} {\bibfnamefont {Y.}~\bibnamefont {Sugita}},\ and\ \bibinfo {author} {\bibfnamefont {H.}~\bibnamefont {Fujiwara}},\ }\href {https://doi.org/10.1103/PhysRevLett.48.1443} {\bibfield  {journal} {\bibinfo  {journal} {Phys. Rev. Lett.}\ }\textbf {\bibinfo {volume} {48}},\ \bibinfo {pages} {1443} (\bibinfo {year} {1982})}\BibitemShut {NoStop}%
\bibitem [{\citenamefont {Webb}\ \emph {et~al.}(1985)\citenamefont {Webb}, \citenamefont {Washburn}, \citenamefont {Umbach},\ and\ \citenamefont {Laibowitz}}]{webb1985observation}%
  \BibitemOpen
  \bibfield  {author} {\bibinfo {author} {\bibfnamefont {R.~A.}\ \bibnamefont {Webb}}, \bibinfo {author} {\bibfnamefont {S.}~\bibnamefont {Washburn}}, \bibinfo {author} {\bibfnamefont {C.~P.}\ \bibnamefont {Umbach}},\ and\ \bibinfo {author} {\bibfnamefont {R.~B.}\ \bibnamefont {Laibowitz}},\ }\href {https://doi.org/10.1103/PhysRevLett.54.2696} {\bibfield  {journal} {\bibinfo  {journal} {Phys. Rev. Lett.}\ }\textbf {\bibinfo {volume} {54}},\ \bibinfo {pages} {2696} (\bibinfo {year} {1985})}\BibitemShut {NoStop}%
\bibitem [{\citenamefont {Matteucci}\ and\ \citenamefont {Pozzi}(1985)}]{matteucci1985new}%
  \BibitemOpen
  \bibfield  {author} {\bibinfo {author} {\bibfnamefont {G.~d.~f.}\ \bibnamefont {Matteucci}}\ and\ \bibinfo {author} {\bibfnamefont {G.}~\bibnamefont {Pozzi}},\ }\href {https://doi.org/10.1103/PhysRevLett.54.2469} {\bibfield  {journal} {\bibinfo  {journal} {Phys. Rev. Lett.}\ }\textbf {\bibinfo {volume} {54}},\ \bibinfo {pages} {2469} (\bibinfo {year} {1985})}\BibitemShut {NoStop}%
\bibitem [{\citenamefont {Tonomura}\ \emph {et~al.}(1986)\citenamefont {Tonomura}, \citenamefont {Osakabe}, \citenamefont {Matsuda}, \citenamefont {Kawasaki}, \citenamefont {Endo}, \citenamefont {Yano},\ and\ \citenamefont {Yamada}}]{tonomura1986evidence}%
  \BibitemOpen
  \bibfield  {author} {\bibinfo {author} {\bibfnamefont {A.}~\bibnamefont {Tonomura}}, \bibinfo {author} {\bibfnamefont {N.}~\bibnamefont {Osakabe}}, \bibinfo {author} {\bibfnamefont {T.}~\bibnamefont {Matsuda}}, \bibinfo {author} {\bibfnamefont {T.}~\bibnamefont {Kawasaki}}, \bibinfo {author} {\bibfnamefont {J.}~\bibnamefont {Endo}}, \bibinfo {author} {\bibfnamefont {S.}~\bibnamefont {Yano}},\ and\ \bibinfo {author} {\bibfnamefont {H.}~\bibnamefont {Yamada}},\ }\href {https://doi.org/10.1103/PhysRevLett.56.792} {\bibfield  {journal} {\bibinfo  {journal} {Phys. Rev. Lett.}\ }\textbf {\bibinfo {volume} {56}},\ \bibinfo {pages} {792} (\bibinfo {year} {1986})}\BibitemShut {NoStop}%
\bibitem [{\citenamefont {Timp}\ \emph {et~al.}(1987)\citenamefont {Timp}, \citenamefont {Chang}, \citenamefont {Cunningham}, \citenamefont {Chang}, \citenamefont {Mankiewich}, \citenamefont {Behringer},\ and\ \citenamefont {Howard}}]{timp1987observation}%
  \BibitemOpen
  \bibfield  {author} {\bibinfo {author} {\bibfnamefont {G.}~\bibnamefont {Timp}}, \bibinfo {author} {\bibfnamefont {A.~M.}\ \bibnamefont {Chang}}, \bibinfo {author} {\bibfnamefont {J.~E.}\ \bibnamefont {Cunningham}}, \bibinfo {author} {\bibfnamefont {T.~Y.}\ \bibnamefont {Chang}}, \bibinfo {author} {\bibfnamefont {P.}~\bibnamefont {Mankiewich}}, \bibinfo {author} {\bibfnamefont {R.}~\bibnamefont {Behringer}},\ and\ \bibinfo {author} {\bibfnamefont {R.~E.}\ \bibnamefont {Howard}},\ }\href {https://doi.org/10.1103/PhysRevLett.58.2814} {\bibfield  {journal} {\bibinfo  {journal} {Phys. Rev. Lett.}\ }\textbf {\bibinfo {volume} {58}},\ \bibinfo {pages} {2814} (\bibinfo {year} {1987})}\BibitemShut {NoStop}%
\bibitem [{\citenamefont {Allman}\ \emph {et~al.}(1992)\citenamefont {Allman}, \citenamefont {Cimmino}, \citenamefont {Klein}, \citenamefont {Opat}, \citenamefont {Kaiser},\ and\ \citenamefont {Werner}}]{allman1992scalar}%
  \BibitemOpen
  \bibfield  {author} {\bibinfo {author} {\bibfnamefont {B.~E.}\ \bibnamefont {Allman}}, \bibinfo {author} {\bibfnamefont {A.}~\bibnamefont {Cimmino}}, \bibinfo {author} {\bibfnamefont {A.~G.}\ \bibnamefont {Klein}}, \bibinfo {author} {\bibfnamefont {G.~I.}\ \bibnamefont {Opat}}, \bibinfo {author} {\bibfnamefont {H.}~\bibnamefont {Kaiser}},\ and\ \bibinfo {author} {\bibfnamefont {S.~A.}\ \bibnamefont {Werner}},\ }\href {https://doi.org/10.1103/PhysRevLett.68.2409} {\bibfield  {journal} {\bibinfo  {journal} {Phys. Rev. Lett.}\ }\textbf {\bibinfo {volume} {68}},\ \bibinfo {pages} {2409} (\bibinfo {year} {1992})}\BibitemShut {NoStop}%
\bibitem [{\citenamefont {Bachtold}\ \emph {et~al.}(1999)\citenamefont {Bachtold}, \citenamefont {Strunk}, \citenamefont {Salvetat}, \citenamefont {Bonard}, \citenamefont {Forr{\'o}}, \citenamefont {Nussbaumer},\ and\ \citenamefont {Sch{\"o}nenberger}}]{bachtold1999aharonov}%
  \BibitemOpen
  \bibfield  {author} {\bibinfo {author} {\bibfnamefont {A.}~\bibnamefont {Bachtold}}, \bibinfo {author} {\bibfnamefont {C.}~\bibnamefont {Strunk}}, \bibinfo {author} {\bibfnamefont {J.-P.}\ \bibnamefont {Salvetat}}, \bibinfo {author} {\bibfnamefont {J.-M.}\ \bibnamefont {Bonard}}, \bibinfo {author} {\bibfnamefont {L.}~\bibnamefont {Forr{\'o}}}, \bibinfo {author} {\bibfnamefont {T.}~\bibnamefont {Nussbaumer}},\ and\ \bibinfo {author} {\bibfnamefont {C.}~\bibnamefont {Sch{\"o}nenberger}},\ }\href {https://www.nature.com/articles/17755} {\bibfield  {journal} {\bibinfo  {journal} {Nature}\ }\textbf {\bibinfo {volume} {397}},\ \bibinfo {pages} {673} (\bibinfo {year} {1999})}\BibitemShut {NoStop}%
\bibitem [{\citenamefont {Haug}\ \emph {et~al.}(2019)\citenamefont {Haug}, \citenamefont {Heimonen}, \citenamefont {Dumke}, \citenamefont {Kwek},\ and\ \citenamefont {Amico}}]{haug2019aharonov}%
  \BibitemOpen
  \bibfield  {author} {\bibinfo {author} {\bibfnamefont {T.}~\bibnamefont {Haug}}, \bibinfo {author} {\bibfnamefont {H.}~\bibnamefont {Heimonen}}, \bibinfo {author} {\bibfnamefont {R.}~\bibnamefont {Dumke}}, \bibinfo {author} {\bibfnamefont {L.-C.}\ \bibnamefont {Kwek}},\ and\ \bibinfo {author} {\bibfnamefont {L.}~\bibnamefont {Amico}},\ }\href {https://doi.org/10.1103/PhysRevA.100.041601} {\bibfield  {journal} {\bibinfo  {journal} {Phys. Rev. A}\ }\textbf {\bibinfo {volume} {100}},\ \bibinfo {pages} {041601} (\bibinfo {year} {2019})}\BibitemShut {NoStop}%
\bibitem [{\citenamefont {Yau}\ \emph {et~al.}(2002)\citenamefont {Yau}, \citenamefont {De~Poortere},\ and\ \citenamefont {Shayegan}}]{yau2002aharonov}%
  \BibitemOpen
  \bibfield  {author} {\bibinfo {author} {\bibfnamefont {J.-B.}\ \bibnamefont {Yau}}, \bibinfo {author} {\bibfnamefont {E.~P.}\ \bibnamefont {De~Poortere}},\ and\ \bibinfo {author} {\bibfnamefont {M.}~\bibnamefont {Shayegan}},\ }\href {https://doi.org/10.1103/PhysRevLett.88.146801} {\bibfield  {journal} {\bibinfo  {journal} {Phys. Rev. Lett.}\ }\textbf {\bibinfo {volume} {88}},\ \bibinfo {pages} {146801} (\bibinfo {year} {2002})}\BibitemShut {NoStop}%
\bibitem [{\citenamefont {Xiao}\ \emph {et~al.}(2010)\citenamefont {Xiao}, \citenamefont {Chang},\ and\ \citenamefont {Niu}}]{xiao2010berry}%
  \BibitemOpen
  \bibfield  {author} {\bibinfo {author} {\bibfnamefont {D.}~\bibnamefont {Xiao}}, \bibinfo {author} {\bibfnamefont {M.-C.}\ \bibnamefont {Chang}},\ and\ \bibinfo {author} {\bibfnamefont {Q.}~\bibnamefont {Niu}},\ }\href {https://doi.org/10.1103/RevModPhys.82.1959} {\bibfield  {journal} {\bibinfo  {journal} {Rev. Mod. Phys.}\ }\textbf {\bibinfo {volume} {82}},\ \bibinfo {pages} {1959} (\bibinfo {year} {2010})}\BibitemShut {NoStop}%
\bibitem [{\citenamefont {Cohen}\ \emph {et~al.}(2019)\citenamefont {Cohen}, \citenamefont {Larocque}, \citenamefont {Bouchard}, \citenamefont {Nejadsattari}, \citenamefont {Gefen},\ and\ \citenamefont {Karimi}}]{cohen2019geometric}%
  \BibitemOpen
  \bibfield  {author} {\bibinfo {author} {\bibfnamefont {E.}~\bibnamefont {Cohen}}, \bibinfo {author} {\bibfnamefont {H.}~\bibnamefont {Larocque}}, \bibinfo {author} {\bibfnamefont {F.}~\bibnamefont {Bouchard}}, \bibinfo {author} {\bibfnamefont {F.}~\bibnamefont {Nejadsattari}}, \bibinfo {author} {\bibfnamefont {Y.}~\bibnamefont {Gefen}},\ and\ \bibinfo {author} {\bibfnamefont {E.}~\bibnamefont {Karimi}},\ }\href {https://www.nature.com/articles/s42254-019-0071-1} {\bibfield  {journal} {\bibinfo  {journal} {Nat. Rev. Phys.}\ }\textbf {\bibinfo {volume} {1}},\ \bibinfo {pages} {437} (\bibinfo {year} {2019})}\BibitemShut {NoStop}%
\bibitem [{\citenamefont {Berry}\ \emph {et~al.}(1980)\citenamefont {Berry}, \citenamefont {Chambers}, \citenamefont {Large}, \citenamefont {Upstill},\ and\ \citenamefont {Walmsley}}]{berry1980wavefront}%
  \BibitemOpen
  \bibfield  {author} {\bibinfo {author} {\bibfnamefont {M.}~\bibnamefont {Berry}}, \bibinfo {author} {\bibfnamefont {R.}~\bibnamefont {Chambers}}, \bibinfo {author} {\bibfnamefont {M.}~\bibnamefont {Large}}, \bibinfo {author} {\bibfnamefont {C.}~\bibnamefont {Upstill}},\ and\ \bibinfo {author} {\bibfnamefont {J.}~\bibnamefont {Walmsley}},\ }\href {https://iopscience.iop.org/article/10.1088/0143-0807/1/3/008} {\bibfield  {journal} {\bibinfo  {journal} {Eur. J. Phys.}\ }\textbf {\bibinfo {volume} {1}},\ \bibinfo {pages} {154} (\bibinfo {year} {1980})}\BibitemShut {NoStop}%
\bibitem [{\citenamefont {Li}\ \emph {et~al.}(2014)\citenamefont {Li}, \citenamefont {Eggleton}, \citenamefont {Fang},\ and\ \citenamefont {Fan}}]{li2014photonic}%
  \BibitemOpen
  \bibfield  {author} {\bibinfo {author} {\bibfnamefont {E.}~\bibnamefont {Li}}, \bibinfo {author} {\bibfnamefont {B.~J.}\ \bibnamefont {Eggleton}}, \bibinfo {author} {\bibfnamefont {K.}~\bibnamefont {Fang}},\ and\ \bibinfo {author} {\bibfnamefont {S.}~\bibnamefont {Fan}},\ }\href {https://www.nature.com/articles/ncomms4225} {\bibfield  {journal} {\bibinfo  {journal} {Nat. Commun.}\ }\textbf {\bibinfo {volume} {5}},\ \bibinfo {pages} {3225} (\bibinfo {year} {2014})}\BibitemShut {NoStop}%
\bibitem [{\citenamefont {Parto}\ \emph {et~al.}(2019)\citenamefont {Parto}, \citenamefont {Lopez-Aviles}, \citenamefont {Antonio-Lopez}, \citenamefont {Khajavikhan}, \citenamefont {Amezcua-Correa},\ and\ \citenamefont {Christodoulides}}]{parto2019observation}%
  \BibitemOpen
  \bibfield  {author} {\bibinfo {author} {\bibfnamefont {M.}~\bibnamefont {Parto}}, \bibinfo {author} {\bibfnamefont {H.}~\bibnamefont {Lopez-Aviles}}, \bibinfo {author} {\bibfnamefont {J.~E.}\ \bibnamefont {Antonio-Lopez}}, \bibinfo {author} {\bibfnamefont {M.}~\bibnamefont {Khajavikhan}}, \bibinfo {author} {\bibfnamefont {R.}~\bibnamefont {Amezcua-Correa}},\ and\ \bibinfo {author} {\bibfnamefont {D.~N.}\ \bibnamefont {Christodoulides}},\ }\href {https://www.science.org/doi/10.1126/sciadv.aau8135} {\bibfield  {journal} {\bibinfo  {journal} {Sci. Adv.}\ }\textbf {\bibinfo {volume} {5}},\ \bibinfo {pages} {eaau8135} (\bibinfo {year} {2019})}\BibitemShut {NoStop}%
\bibitem [{\citenamefont {Vaidman}(2012)}]{vaidman2012role}%
  \BibitemOpen
  \bibfield  {author} {\bibinfo {author} {\bibfnamefont {L.}~\bibnamefont {Vaidman}},\ }\href {https://journals.aps.org/pra/abstract/10.1103/PhysRevA.86.040101} {\bibfield  {journal} {\bibinfo  {journal} {Phys. Rev. A}\ }\textbf {\bibinfo {volume} {86}},\ \bibinfo {pages} {040101} (\bibinfo {year} {2012})}\BibitemShut {NoStop}%
\bibitem [{\citenamefont {Aharonov}\ \emph {et~al.}(2015)\citenamefont {Aharonov}, \citenamefont {Cohen},\ and\ \citenamefont {Rohrlich}}]{aharonov2015comment}%
  \BibitemOpen
  \bibfield  {author} {\bibinfo {author} {\bibfnamefont {Y.}~\bibnamefont {Aharonov}}, \bibinfo {author} {\bibfnamefont {E.}~\bibnamefont {Cohen}},\ and\ \bibinfo {author} {\bibfnamefont {D.}~\bibnamefont {Rohrlich}},\ }\href {https://journals.aps.org/pra/abstract/10.1103/PhysRevA.92.026101} {\bibfield  {journal} {\bibinfo  {journal} {Phys. Rev. A}\ }\textbf {\bibinfo {volume} {92}},\ \bibinfo {pages} {026101} (\bibinfo {year} {2015})}\BibitemShut {NoStop}%
\bibitem [{\citenamefont {Vaidman}(2015)}]{vaidman2015reply}%
  \BibitemOpen
  \bibfield  {author} {\bibinfo {author} {\bibfnamefont {L.}~\bibnamefont {Vaidman}},\ }\href {https://journals.aps.org/pra/abstract/10.1103/PhysRevA.92.026102} {\bibfield  {journal} {\bibinfo  {journal} {Phys. Rev. A}\ }\textbf {\bibinfo {volume} {92}},\ \bibinfo {pages} {026102} (\bibinfo {year} {2015})}\BibitemShut {NoStop}%
\bibitem [{\citenamefont {Aharonov}\ \emph {et~al.}(2016)\citenamefont {Aharonov}, \citenamefont {Cohen},\ and\ \citenamefont {Rohrlich}}]{aharonov2016nonlocality}%
  \BibitemOpen
  \bibfield  {author} {\bibinfo {author} {\bibfnamefont {Y.}~\bibnamefont {Aharonov}}, \bibinfo {author} {\bibfnamefont {E.}~\bibnamefont {Cohen}},\ and\ \bibinfo {author} {\bibfnamefont {D.}~\bibnamefont {Rohrlich}},\ }\href {https://journals.aps.org/pra/abstract/10.1103/PhysRevA.93.042110} {\bibfield  {journal} {\bibinfo  {journal} {Phys. Rev. A}\ }\textbf {\bibinfo {volume} {93}},\ \bibinfo {pages} {042110} (\bibinfo {year} {2016})}\BibitemShut {NoStop}%
\bibitem [{\citenamefont {Li}\ \emph {et~al.}(2022)\citenamefont {Li}, \citenamefont {Hansson},\ and\ \citenamefont {Ku}}]{li2022gauge}%
  \BibitemOpen
  \bibfield  {author} {\bibinfo {author} {\bibfnamefont {X.}~\bibnamefont {Li}}, \bibinfo {author} {\bibfnamefont {T.~H.}\ \bibnamefont {Hansson}},\ and\ \bibinfo {author} {\bibfnamefont {W.}~\bibnamefont {Ku}},\ }\href {https://doi.org/10.1103/PhysRevA.106.032217} {\bibfield  {journal} {\bibinfo  {journal} {Phys. Rev. A}\ }\textbf {\bibinfo {volume} {106}},\ \bibinfo {pages} {032217} (\bibinfo {year} {2022})}\BibitemShut {NoStop}%
\bibitem [{\citenamefont {Paiva}\ \emph {et~al.}(2023)\citenamefont {Paiva}, \citenamefont {Dieguez}, \citenamefont {Angelo},\ and\ \citenamefont {Cohen}}]{paiva2023coherence}%
  \BibitemOpen
  \bibfield  {author} {\bibinfo {author} {\bibfnamefont {I.~L.}\ \bibnamefont {Paiva}}, \bibinfo {author} {\bibfnamefont {P.~R.}\ \bibnamefont {Dieguez}}, \bibinfo {author} {\bibfnamefont {R.~M.}\ \bibnamefont {Angelo}},\ and\ \bibinfo {author} {\bibfnamefont {E.}~\bibnamefont {Cohen}},\ }\href {https://journals.aps.org/pra/abstract/10.1103/PhysRevA.107.032213} {\bibfield  {journal} {\bibinfo  {journal} {Phys. Rev. A}\ }\textbf {\bibinfo {volume} {107}},\ \bibinfo {pages} {032213} (\bibinfo {year} {2023})}\BibitemShut {NoStop}%
\bibitem [{\citenamefont {Chang}\ and\ \citenamefont {Niu}(1995)}]{changniu1995}%
  \BibitemOpen
  \bibfield  {author} {\bibinfo {author} {\bibfnamefont {M.-C.}\ \bibnamefont {Chang}}\ and\ \bibinfo {author} {\bibfnamefont {Q.}~\bibnamefont {Niu}},\ }\href {https://doi.org/10.1103/PhysRevLett.75.1348} {\bibfield  {journal} {\bibinfo  {journal} {Phys. Rev. Lett.}\ }\textbf {\bibinfo {volume} {75}},\ \bibinfo {pages} {1348} (\bibinfo {year} {1995})}\BibitemShut {NoStop}%
\bibitem [{\citenamefont {Chang}\ and\ \citenamefont {Niu}(1996)}]{chang1996berry}%
  \BibitemOpen
  \bibfield  {author} {\bibinfo {author} {\bibfnamefont {M.-C.}\ \bibnamefont {Chang}}\ and\ \bibinfo {author} {\bibfnamefont {Q.}~\bibnamefont {Niu}},\ }\href {https://journals.aps.org/prb/abstract/10.1103/PhysRevB.53.7010} {\bibfield  {journal} {\bibinfo  {journal} {Phys. Rev. B}\ }\textbf {\bibinfo {volume} {53}},\ \bibinfo {pages} {7010} (\bibinfo {year} {1996})}\BibitemShut {NoStop}%
\bibitem [{\citenamefont {Price}\ \emph {et~al.}(2016)\citenamefont {Price}, \citenamefont {Zilberberg}, \citenamefont {Ozawa}, \citenamefont {Carusotto},\ and\ \citenamefont {Goldman}}]{price2016measurement}%
  \BibitemOpen
  \bibfield  {author} {\bibinfo {author} {\bibfnamefont {H.~M.}\ \bibnamefont {Price}}, \bibinfo {author} {\bibfnamefont {O.}~\bibnamefont {Zilberberg}}, \bibinfo {author} {\bibfnamefont {T.}~\bibnamefont {Ozawa}}, \bibinfo {author} {\bibfnamefont {I.}~\bibnamefont {Carusotto}},\ and\ \bibinfo {author} {\bibfnamefont {N.}~\bibnamefont {Goldman}},\ }\href {https://journals.aps.org/prb/abstract/10.1103/PhysRevB.93.245113} {\bibfield  {journal} {\bibinfo  {journal} {Phys. Rev. B}\ }\textbf {\bibinfo {volume} {93}},\ \bibinfo {pages} {245113} (\bibinfo {year} {2016})}\BibitemShut {NoStop}%
\bibitem [{\citenamefont {Wimmer}\ \emph {et~al.}(2017)\citenamefont {Wimmer}, \citenamefont {Price}, \citenamefont {Carusotto},\ and\ \citenamefont {Peschel}}]{wimmer2017experimental}%
  \BibitemOpen
  \bibfield  {author} {\bibinfo {author} {\bibfnamefont {M.}~\bibnamefont {Wimmer}}, \bibinfo {author} {\bibfnamefont {H.~M.}\ \bibnamefont {Price}}, \bibinfo {author} {\bibfnamefont {I.}~\bibnamefont {Carusotto}},\ and\ \bibinfo {author} {\bibfnamefont {U.}~\bibnamefont {Peschel}},\ }\href {https://www.nature.com/articles/nphys4050} {\bibfield  {journal} {\bibinfo  {journal} {Nat. Phys.}\ }\textbf {\bibinfo {volume} {13}},\ \bibinfo {pages} {545} (\bibinfo {year} {2017})}\BibitemShut {NoStop}%
\bibitem [{\citenamefont {D’Errico}\ \emph {et~al.}(2020)\citenamefont {D’Errico}, \citenamefont {Cardano}, \citenamefont {Maffei}, \citenamefont {Dauphin}, \citenamefont {Barboza}, \citenamefont {Esposito}, \citenamefont {Piccirillo}, \citenamefont {Lewenstein}, \citenamefont {Massignan},\ and\ \citenamefont {Marrucci}}]{d2020two}%
  \BibitemOpen
  \bibfield  {author} {\bibinfo {author} {\bibfnamefont {A.}~\bibnamefont {D’Errico}}, \bibinfo {author} {\bibfnamefont {F.}~\bibnamefont {Cardano}}, \bibinfo {author} {\bibfnamefont {M.}~\bibnamefont {Maffei}}, \bibinfo {author} {\bibfnamefont {A.}~\bibnamefont {Dauphin}}, \bibinfo {author} {\bibfnamefont {R.}~\bibnamefont {Barboza}}, \bibinfo {author} {\bibfnamefont {C.}~\bibnamefont {Esposito}}, \bibinfo {author} {\bibfnamefont {B.}~\bibnamefont {Piccirillo}}, \bibinfo {author} {\bibfnamefont {M.}~\bibnamefont {Lewenstein}}, \bibinfo {author} {\bibfnamefont {P.}~\bibnamefont {Massignan}},\ and\ \bibinfo {author} {\bibfnamefont {L.}~\bibnamefont {Marrucci}},\ }\href {https://doi.org/10.1364/OPTICA.365028} {\bibfield  {journal} {\bibinfo  {journal} {Optica}\ }\textbf {\bibinfo {volume} {7}},\ \bibinfo {pages} {108} (\bibinfo {year} {2020})}\BibitemShut {NoStop}%
\bibitem [{\citenamefont {Bernevig}(2013)}]{bernevig2013topological}%
  \BibitemOpen
  \bibfield  {author} {\bibinfo {author} {\bibfnamefont {B.~A.}\ \bibnamefont {Bernevig}},\ }\href {https://press.princeton.edu/books/hardcover/9780691151755/topological-insulators-and-topological-superconductors} {\emph {\bibinfo {title} {Topological insulators and topological superconductors}}}\ (\bibinfo  {publisher} {Princeton University Press},\ \bibinfo {year} {2013})\BibitemShut {NoStop}%
\bibitem [{\citenamefont {Asb{\'o}th}\ \emph {et~al.}(2016)\citenamefont {Asb{\'o}th}, \citenamefont {Oroszl{\'a}ny}, \citenamefont {P{\'a}lyi}, \citenamefont {Asb{\'o}th}, \citenamefont {Oroszl{\'a}ny},\ and\ \citenamefont {P{\'a}lyi}}]{asboth2016berry}%
  \BibitemOpen
  \bibfield  {author} {\bibinfo {author} {\bibfnamefont {J.~K.}\ \bibnamefont {Asb{\'o}th}}, \bibinfo {author} {\bibfnamefont {L.}~\bibnamefont {Oroszl{\'a}ny}}, \bibinfo {author} {\bibfnamefont {A.}~\bibnamefont {P{\'a}lyi}}, \bibinfo {author} {\bibfnamefont {J.~K.}\ \bibnamefont {Asb{\'o}th}}, \bibinfo {author} {\bibfnamefont {L.}~\bibnamefont {Oroszl{\'a}ny}},\ and\ \bibinfo {author} {\bibfnamefont {A.}~\bibnamefont {P{\'a}lyi}},\ }\href {https://link.springer.com/book/10.1007/978-3-319-25607-8} {\emph {\bibinfo {title} {A Short Course on Topological Insulators}}}\ (\bibinfo  {publisher} {Springer},\ \bibinfo {year} {2016})\BibitemShut {NoStop}%
\bibitem [{\citenamefont {Duca}\ \emph {et~al.}(2015)\citenamefont {Duca}, \citenamefont {Li}, \citenamefont {Reitter}, \citenamefont {Bloch}, \citenamefont {Schleier-Smith},\ and\ \citenamefont {Schneider}}]{duca2015aharonov}%
  \BibitemOpen
  \bibfield  {author} {\bibinfo {author} {\bibfnamefont {L.}~\bibnamefont {Duca}}, \bibinfo {author} {\bibfnamefont {T.}~\bibnamefont {Li}}, \bibinfo {author} {\bibfnamefont {M.}~\bibnamefont {Reitter}}, \bibinfo {author} {\bibfnamefont {I.}~\bibnamefont {Bloch}}, \bibinfo {author} {\bibfnamefont {M.}~\bibnamefont {Schleier-Smith}},\ and\ \bibinfo {author} {\bibfnamefont {U.}~\bibnamefont {Schneider}},\ }\href {https://www.science.org/doi/10.1126/science.1259052} {\bibfield  {journal} {\bibinfo  {journal} {Science}\ }\textbf {\bibinfo {volume} {347}},\ \bibinfo {pages} {288} (\bibinfo {year} {2015})}\BibitemShut {NoStop}%
\bibitem [{\citenamefont {Su}\ \emph {et~al.}(1980)\citenamefont {Su}, \citenamefont {Schrieffer},\ and\ \citenamefont {Heeger}}]{su1980soliton}%
  \BibitemOpen
  \bibfield  {author} {\bibinfo {author} {\bibfnamefont {W.-P.}\ \bibnamefont {Su}}, \bibinfo {author} {\bibfnamefont {J.}~\bibnamefont {Schrieffer}},\ and\ \bibinfo {author} {\bibfnamefont {A.}~\bibnamefont {Heeger}},\ }\href {https://doi.org/10.1103/PhysRevB.22.2099} {\bibfield  {journal} {\bibinfo  {journal} {Phys. Rev. B}\ }\textbf {\bibinfo {volume} {22}},\ \bibinfo {pages} {2099} (\bibinfo {year} {1980})}\BibitemShut {NoStop}%
\bibitem [{\citenamefont {Castro~Neto}\ \emph {et~al.}(2009)\citenamefont {Castro~Neto}, \citenamefont {Guinea}, \citenamefont {Peres}, \citenamefont {Novoselov},\ and\ \citenamefont {Geim}}]{RevModPhys.81.109}%
  \BibitemOpen
  \bibfield  {author} {\bibinfo {author} {\bibfnamefont {A.~H.}\ \bibnamefont {Castro~Neto}}, \bibinfo {author} {\bibfnamefont {F.}~\bibnamefont {Guinea}}, \bibinfo {author} {\bibfnamefont {N.~M.~R.}\ \bibnamefont {Peres}}, \bibinfo {author} {\bibfnamefont {K.~S.}\ \bibnamefont {Novoselov}},\ and\ \bibinfo {author} {\bibfnamefont {A.~K.}\ \bibnamefont {Geim}},\ }\href {https://doi.org/10.1103/RevModPhys.81.109} {\bibfield  {journal} {\bibinfo  {journal} {Rev. Mod. Phys.}\ }\textbf {\bibinfo {volume} {81}},\ \bibinfo {pages} {109} (\bibinfo {year} {2009})}\BibitemShut {NoStop}%
\bibitem [{\citenamefont {Cardano}\ \emph {et~al.}(2017)\citenamefont {Cardano}, \citenamefont {D’Errico}, \citenamefont {Dauphin}, \citenamefont {Maffei}, \citenamefont {Piccirillo}, \citenamefont {de~Lisio}, \citenamefont {De~Filippis}, \citenamefont {Cataudella}, \citenamefont {Santamato},\ and\ \citenamefont {Marrucci}}]{cardano2017detection}%
  \BibitemOpen
  \bibfield  {author} {\bibinfo {author} {\bibfnamefont {F.}~\bibnamefont {Cardano}}, \bibinfo {author} {\bibfnamefont {A.}~\bibnamefont {D’Errico}}, \bibinfo {author} {\bibfnamefont {A.}~\bibnamefont {Dauphin}}, \bibinfo {author} {\bibfnamefont {M.}~\bibnamefont {Maffei}}, \bibinfo {author} {\bibfnamefont {B.}~\bibnamefont {Piccirillo}}, \bibinfo {author} {\bibfnamefont {C.}~\bibnamefont {de~Lisio}}, \bibinfo {author} {\bibfnamefont {G.}~\bibnamefont {De~Filippis}}, \bibinfo {author} {\bibfnamefont {V.}~\bibnamefont {Cataudella}}, \bibinfo {author} {\bibfnamefont {E.}~\bibnamefont {Santamato}},\ and\ \bibinfo {author} {\bibfnamefont {L.}~\bibnamefont {Marrucci}},\ }\href {https://www.nature.com/articles/ncomms15516} {\bibfield  {journal} {\bibinfo  {journal} {Nat. Commun.}\ }\textbf {\bibinfo {volume} {8}},\ \bibinfo {pages} {15516} (\bibinfo {year} {2017})}\BibitemShut {NoStop}%
\bibitem [{\citenamefont {Maffei}\ \emph {et~al.}(2018)\citenamefont {Maffei}, \citenamefont {Dauphin}, \citenamefont {Cardano}, \citenamefont {Lewenstein},\ and\ \citenamefont {Massignan}}]{maffei2018topological}%
  \BibitemOpen
  \bibfield  {author} {\bibinfo {author} {\bibfnamefont {M.}~\bibnamefont {Maffei}}, \bibinfo {author} {\bibfnamefont {A.}~\bibnamefont {Dauphin}}, \bibinfo {author} {\bibfnamefont {F.}~\bibnamefont {Cardano}}, \bibinfo {author} {\bibfnamefont {M.}~\bibnamefont {Lewenstein}},\ and\ \bibinfo {author} {\bibfnamefont {P.}~\bibnamefont {Massignan}},\ }\href {https://iopscience.iop.org/article/10.1088/1367-2630/aa9d4c} {\bibfield  {journal} {\bibinfo  {journal} {New J. Phys.}\ }\textbf {\bibinfo {volume} {20}},\ \bibinfo {pages} {013023} (\bibinfo {year} {2018})}\BibitemShut {NoStop}%
\bibitem [{\citenamefont {D'Errico}\ \emph {et~al.}(2020)\citenamefont {D'Errico}, \citenamefont {Di~Colandrea}, \citenamefont {Barboza}, \citenamefont {Dauphin}, \citenamefont {Lewenstein}, \citenamefont {Massignan}, \citenamefont {Marrucci},\ and\ \citenamefont {Cardano}}]{d2020bulk}%
  \BibitemOpen
  \bibfield  {author} {\bibinfo {author} {\bibfnamefont {A.}~\bibnamefont {D'Errico}}, \bibinfo {author} {\bibfnamefont {F.}~\bibnamefont {Di~Colandrea}}, \bibinfo {author} {\bibfnamefont {R.}~\bibnamefont {Barboza}}, \bibinfo {author} {\bibfnamefont {A.}~\bibnamefont {Dauphin}}, \bibinfo {author} {\bibfnamefont {M.}~\bibnamefont {Lewenstein}}, \bibinfo {author} {\bibfnamefont {P.}~\bibnamefont {Massignan}}, \bibinfo {author} {\bibfnamefont {L.}~\bibnamefont {Marrucci}},\ and\ \bibinfo {author} {\bibfnamefont {F.}~\bibnamefont {Cardano}},\ }\href {https://doi.org/10.1103/PhysRevResearch.2.023119} {\bibfield  {journal} {\bibinfo  {journal} {Phys. Rev. Res.}\ }\textbf {\bibinfo {volume} {2}},\ \bibinfo {pages} {023119} (\bibinfo {year} {2020})}\BibitemShut {NoStop}%
\bibitem [{\citenamefont {Meier}\ \emph {et~al.}(2018)\citenamefont {Meier}, \citenamefont {An}, \citenamefont {Dauphin}, \citenamefont {Maffei}, \citenamefont {Massignan}, \citenamefont {Hughes},\ and\ \citenamefont {Gadway}}]{meier2018observation}%
  \BibitemOpen
  \bibfield  {author} {\bibinfo {author} {\bibfnamefont {E.~J.}\ \bibnamefont {Meier}}, \bibinfo {author} {\bibfnamefont {F.~A.}\ \bibnamefont {An}}, \bibinfo {author} {\bibfnamefont {A.}~\bibnamefont {Dauphin}}, \bibinfo {author} {\bibfnamefont {M.}~\bibnamefont {Maffei}}, \bibinfo {author} {\bibfnamefont {P.}~\bibnamefont {Massignan}}, \bibinfo {author} {\bibfnamefont {T.~L.}\ \bibnamefont {Hughes}},\ and\ \bibinfo {author} {\bibfnamefont {B.}~\bibnamefont {Gadway}},\ }\href {https://www.science.org/doi/10.1126/science.aat3406} {\bibfield  {journal} {\bibinfo  {journal} {Science}\ }\textbf {\bibinfo {volume} {362}},\ \bibinfo {pages} {929} (\bibinfo {year} {2018})}\BibitemShut {NoStop}%
\bibitem [{\citenamefont {Haller}\ \emph {et~al.}(2020)\citenamefont {Haller}, \citenamefont {Massignan},\ and\ \citenamefont {Rizzi}}]{haller2020detecting}%
  \BibitemOpen
  \bibfield  {author} {\bibinfo {author} {\bibfnamefont {A.}~\bibnamefont {Haller}}, \bibinfo {author} {\bibfnamefont {P.}~\bibnamefont {Massignan}},\ and\ \bibinfo {author} {\bibfnamefont {M.}~\bibnamefont {Rizzi}},\ }\href {https://doi.org/10.1103/PhysRevResearch.2.033200} {\bibfield  {journal} {\bibinfo  {journal} {Phys. Rev. Res.}\ }\textbf {\bibinfo {volume} {2}},\ \bibinfo {pages} {033200} (\bibinfo {year} {2020})}\BibitemShut {NoStop}%
\bibitem [{\citenamefont {St-Jean}\ \emph {et~al.}(2021)\citenamefont {St-Jean}, \citenamefont {Dauphin}, \citenamefont {Massignan}, \citenamefont {Real}, \citenamefont {Jamadi}, \citenamefont {Milicevic}, \citenamefont {Lema\^{\i}tre}, \citenamefont {Harouri}, \citenamefont {Le~Gratiet}, \citenamefont {Sagnes}, \citenamefont {Ravets}, \citenamefont {Bloch},\ and\ \citenamefont {Amo}}]{st2021measuring}%
  \BibitemOpen
  \bibfield  {author} {\bibinfo {author} {\bibfnamefont {P.}~\bibnamefont {St-Jean}}, \bibinfo {author} {\bibfnamefont {A.}~\bibnamefont {Dauphin}}, \bibinfo {author} {\bibfnamefont {P.}~\bibnamefont {Massignan}}, \bibinfo {author} {\bibfnamefont {B.}~\bibnamefont {Real}}, \bibinfo {author} {\bibfnamefont {O.}~\bibnamefont {Jamadi}}, \bibinfo {author} {\bibfnamefont {M.}~\bibnamefont {Milicevic}}, \bibinfo {author} {\bibfnamefont {A.}~\bibnamefont {Lema\^{\i}tre}}, \bibinfo {author} {\bibfnamefont {A.}~\bibnamefont {Harouri}}, \bibinfo {author} {\bibfnamefont {L.}~\bibnamefont {Le~Gratiet}}, \bibinfo {author} {\bibfnamefont {I.}~\bibnamefont {Sagnes}}, \bibinfo {author} {\bibfnamefont {S.}~\bibnamefont {Ravets}}, \bibinfo {author} {\bibfnamefont {J.}~\bibnamefont {Bloch}},\ and\ \bibinfo {author} {\bibfnamefont {A.}~\bibnamefont {Amo}},\ }\href {https://doi.org/10.1103/PhysRevLett.126.127403} {\bibfield  {journal} {\bibinfo  {journal} {Phys. Rev. Lett.}\ }\textbf {\bibinfo {volume} {126}},\ \bibinfo {pages}
  {127403} (\bibinfo {year} {2021})}\BibitemShut {NoStop}%
\bibitem [{\citenamefont {Bliokh}\ \emph {et~al.}(2019)\citenamefont {Bliokh}, \citenamefont {Alonso},\ and\ \citenamefont {Dennis}}]{bliokh2019geometric}%
  \BibitemOpen
  \bibfield  {author} {\bibinfo {author} {\bibfnamefont {K.~Y.}\ \bibnamefont {Bliokh}}, \bibinfo {author} {\bibfnamefont {M.~A.}\ \bibnamefont {Alonso}},\ and\ \bibinfo {author} {\bibfnamefont {M.~R.}\ \bibnamefont {Dennis}},\ }\href {https://doi.org/10.1088/1361-6633/ab4415} {\bibfield  {journal} {\bibinfo  {journal} {Rep. Prog. Phys.}\ }\textbf {\bibinfo {volume} {82}},\ \bibinfo {pages} {122401} (\bibinfo {year} {2019})}\BibitemShut {NoStop}%
\bibitem [{\citenamefont {D’Errico}\ \emph {et~al.}(2021)\citenamefont {D’Errico}, \citenamefont {Barboza}, \citenamefont {Tudor}, \citenamefont {Dauphin}, \citenamefont {Massignan}, \citenamefont {Marrucci},\ and\ \citenamefont {Cardano}}]{d2021bloch}%
  \BibitemOpen
  \bibfield  {author} {\bibinfo {author} {\bibfnamefont {A.}~\bibnamefont {D’Errico}}, \bibinfo {author} {\bibfnamefont {R.}~\bibnamefont {Barboza}}, \bibinfo {author} {\bibfnamefont {R.}~\bibnamefont {Tudor}}, \bibinfo {author} {\bibfnamefont {A.}~\bibnamefont {Dauphin}}, \bibinfo {author} {\bibfnamefont {P.}~\bibnamefont {Massignan}}, \bibinfo {author} {\bibfnamefont {L.}~\bibnamefont {Marrucci}},\ and\ \bibinfo {author} {\bibfnamefont {F.}~\bibnamefont {Cardano}},\ }\href {https://doi.org/10.1063/5.0037327} {\bibfield  {journal} {\bibinfo  {journal} {APL Photonics}\ }\textbf {\bibinfo {volume} {6}} (\bibinfo {year} {2021})}\BibitemShut {NoStop}%
\bibitem [{\citenamefont {Esposito}\ \emph {et~al.}(2022)\citenamefont {Esposito}, \citenamefont {Barros}, \citenamefont {Dur{\'a}n~Hern{\'a}ndez}, \citenamefont {Carvacho}, \citenamefont {Di~Colandrea}, \citenamefont {Barboza}, \citenamefont {Cardano}, \citenamefont {Spagnolo}, \citenamefont {Marrucci},\ and\ \citenamefont {Sciarrino}}]{esposito2022quantum}%
  \BibitemOpen
  \bibfield  {author} {\bibinfo {author} {\bibfnamefont {C.}~\bibnamefont {Esposito}}, \bibinfo {author} {\bibfnamefont {M.~R.}\ \bibnamefont {Barros}}, \bibinfo {author} {\bibfnamefont {A.}~\bibnamefont {Dur{\'a}n~Hern{\'a}ndez}}, \bibinfo {author} {\bibfnamefont {G.}~\bibnamefont {Carvacho}}, \bibinfo {author} {\bibfnamefont {F.}~\bibnamefont {Di~Colandrea}}, \bibinfo {author} {\bibfnamefont {R.}~\bibnamefont {Barboza}}, \bibinfo {author} {\bibfnamefont {F.}~\bibnamefont {Cardano}}, \bibinfo {author} {\bibfnamefont {N.}~\bibnamefont {Spagnolo}}, \bibinfo {author} {\bibfnamefont {L.}~\bibnamefont {Marrucci}},\ and\ \bibinfo {author} {\bibfnamefont {F.}~\bibnamefont {Sciarrino}},\ }\href {https://www.nature.com/articles/s41534-022-00544-0} {\bibfield  {journal} {\bibinfo  {journal} {Npj Quantum Inf.}\ }\textbf {\bibinfo {volume} {8}},\ \bibinfo {pages} {34} (\bibinfo {year} {2022})}\BibitemShut {NoStop}%
\bibitem [{\citenamefont {Di~Colandrea}\ \emph {et~al.}(2023)\citenamefont {Di~Colandrea}, \citenamefont {Babazadeh}, \citenamefont {Dauphin}, \citenamefont {Massignan}, \citenamefont {Marrucci},\ and\ \citenamefont {Cardano}}]{di2023ultra}%
  \BibitemOpen
  \bibfield  {author} {\bibinfo {author} {\bibfnamefont {F.}~\bibnamefont {Di~Colandrea}}, \bibinfo {author} {\bibfnamefont {A.}~\bibnamefont {Babazadeh}}, \bibinfo {author} {\bibfnamefont {A.}~\bibnamefont {Dauphin}}, \bibinfo {author} {\bibfnamefont {P.}~\bibnamefont {Massignan}}, \bibinfo {author} {\bibfnamefont {L.}~\bibnamefont {Marrucci}},\ and\ \bibinfo {author} {\bibfnamefont {F.}~\bibnamefont {Cardano}},\ }\href {https://doi.org/10.1364/OPTICA.474542} {\bibfield  {journal} {\bibinfo  {journal} {Optica}\ }\textbf {\bibinfo {volume} {10}},\ \bibinfo {pages} {324} (\bibinfo {year} {2023})}\BibitemShut {NoStop}%
\end{thebibliography}%
\vspace{0.5 cm}
\noindent\textbf{Author contributions:} AD conceived the idea and developed the theory. FDC, with contributions from ND and AD, performed the simulations. FDC and ND fabricated the plates and performed the experiment and data analysis. FC and EK supervised the project. All authors contributed to the writing of the manuscript.

\clearpage
\onecolumngrid
\renewcommand{\figurename}{\textbf{Figure}}
\setcounter{figure}{0} \renewcommand{\thefigure}{\textbf{S{\arabic{figure}}}}
\setcounter{table}{0} \renewcommand{\thetable}{S\arabic{table}}
\setcounter{section}{0} \renewcommand{\thesection}{S\arabic{section}}
\setcounter{equation}{0} \renewcommand{\theequation}{S\arabic{equation}}
\onecolumngrid

\begin{center}
{\Large Supplementary Material for: \\ Observation of the Berry connection in chiral lattice systems}
\end{center}
\vspace{1 EM}
\subsection*{I. Proof of the main result.}

From Eq.~\eqref{eq:MCD}, using Eq.~\eqref{chiralU} and the momentum representation of the position operator, ${\bra{\mathbf{q}}\hat{x}_i\ket{\mathbf{q}'}=i\delta(\mathbf{q}-\mathbf{q}')\partial_{q_i}}$, we obtain
\begin{equation}
\begin{split}
\mathcal{C}_i(t)=&\,2i\int_\text{BZ}\frac{d^D q}{(2\pi)^D}\abs{G_{w,\mathbf{q}_0}}^2\bra{\phi_0}\mathcal{U}^{-t}\hat{\Gamma}\partial_{q_i}\mathcal{U}^t\ket{\phi_0}\cr &+2i\int_\text{BZ}\frac{d^D q}{(2\pi)^D}\bra{\phi_0}\mathcal{U}^{-t}\hat{\Gamma}\mathcal{U}^t\ket{\phi_0}\frac{\partial_{q_i}\abs{G_{w,\mathbf{q}_0}}^2}{2},
\end{split}
\end{equation}
where we suppressed the dependency on the quasi-momentum to simplify the notation.
Using ${\mathcal{U}^t=\cos(Et) \sigma_0-i\sin(Et)\mathbf{n}\cdot\hat{\boldsymbol{\sigma}}}$, where $\sigma_0$ is the identity operator, and ${\mathcal{U}^{-t}\Gamma=\Gamma\mathcal{U}^t}$, 
we obtain
\begin{align}
\mathcal{U}^{-t}\hat{\Gamma}\partial_{q_i}\mathcal{U}^t=&\frac{1}{2}\hat{\Gamma}\partial_{q_i}(\cos(2tE)-i\sin(2tE)\mathbf{n}\cdot\hat{\boldsymbol{\sigma}})\cr&-i\sin^2(tE)\left(\mathbf{n}\cross \partial_{q_i}\mathbf{n}\right)\cdot \mathbf{v}_{\Gamma}\cr=&\frac{1}{2}\hat{\Gamma}\partial_{q_i}\mathcal{U}^{2t}-i\sin^2(tE)2\mathcal{A}_i.
\end{align}
In the last equality, we used ${\mathcal{A}_i=i\bra{\mathbf{n}(\mathbf{q})}\partial_{q_i}\ket{\mathbf{n}(\mathbf{q})}=(\mathbf{n}\cross \partial_{q_i}\mathbf{n})\cdot \mathbf{v}_{\Gamma}/2}$, with ${\ket{\mathbf{n}(\mathbf{q})}=(e^{-i\phi(\mathbf{q})}\ket{\uparrow}+\ket{\downarrow})/\sqrt{2}}$, and ${\phi:=\arctan{(n_y/n_x)}}$. 
Thus,
\begin{equation}
\begin{split}
\mathcal{C}_i(t)=&\,2\int_\text{BZ}\frac{d^D q}{(2\pi)^D}\abs{G_{w,\mathbf{q}_0}}^2\sin^2(tE)2\mathcal{A}_i\cr &+2i\int_\text{BZ}\frac{d^D q}{(2\pi)^D}\abs{G_{w,\mathbf{q}_0}}^2\frac{1}{2}\bra{\phi_0}\hat{\Gamma}\partial_{q_i}\mathcal{U}^{2t}\ket{\phi_0}\cr&+2i\int_\text{BZ}\frac{d^D q}{(2\pi)^D}\bra{\phi_0}\hat{\Gamma}\mathcal{U}^{2t}\ket{\phi_0}\frac{\partial_{q_i}\abs{G_{w,\mathbf{q}_0}}^2}{2}.
\end{split}
\end{equation}
Integration by parts shows that the last two terms cancel each other:
\begin{align}
\mathcal{C}_i(t)=&\,2\int_\text{BZ}\frac{d^D q}{(2\pi)^D}\abs{G_{w,\mathbf{q}_0}}^2\sin^2(tE)2\mathcal{A}_i\cr=&\int_\text{BZ}\frac{d^D q}{(2\pi)^D}\abs{G_{w,\mathbf{q}_0}}^2 2\mathcal{A}_i\cr&-\int_\text{BZ}\frac{d^D q}{(2\pi)^D}\abs{G_{w,\mathbf{q}_0}}^2\cos(2Et)2\mathcal{A}_i.
\label{eq:MCDtotsupp}
\end{align}
Using the stationary phase approximation, it can be shown that the last integral gives a contribution that oscillates in time and generally decreases in amplitude as $\sim 1/\sqrt{t}$, from which the final result of Eq.~\eqref{eq:Ci} is derived. Note that, in the case of flat bands $E(\mathbf{q})=\text{constant}$, the asymptotic result is exact apart from a $2\sin^2(tE)$ multiplicative factor.

As discussed in the main text, if ${\abs{G_{w,\mathbf{q}_0}}^2=g_w(\mathbf{q}-\mathbf{q}_0)}$, the Berry connection can be extracted from the time average of the deconvolved MCD.

\subsection*{II. Details on numerical simulations}

\textbf{SSH Hamiltonian.} The SSH model~\cite{su1980soliton}  describes a composite 1D lattice with two sites per unit cell. The lattice Hamiltonian is 
\begin{equation}
H=\sum_{x} a\ket{x,B}\bra{x,A}+b\ket{x+1,A}\bra{x,B}+\text{h.c.},
\end{equation}
where $x\in \mathbb{Z} $ labels the lattice sites, the two sublattices are labeled as $A$ and $B$, and $\text{h.c.}$ denotes the Hermitian conjugate. The coefficients $a$ and $b$ are the intracell and intercell hopping amplitudes, respectively. 
The Bloch theorem allows diagonalizing $H$ as ${H=\int(dq/2\pi) \,\mathcal{H}(q)\otimes \ketbra{q}}$, with ${\mathcal{H}(q)=E(q)\mathbf{n}(q)\cdot\hat{\boldsymbol{\sigma}}}$, where \begin{align}E(q)&=\sqrt{a^2+2 a b \cos (q)+b^2},\cr n_x(q)&=\frac{a+b \cos (q)}{E(q)},\cr n_y(q)&=\frac{b \sin (q)}{E(q)},\cr n_z(q)&=0.\end{align} The chiral operator is $\Gamma=\sigma_z$.

\textbf{Graphene Hamiltonian.} A simple nearest-neighbor tight-binding model for graphene, which only includes the two energy bands near the Fermi energy, gives the following Bloch Hamiltonian  $\mathcal{H}$~\cite{RevModPhys.81.109}: 
\begin{align}
\mathcal{H}(q_x,q_y)=&-\tau\sigma_x \biggl[\cos \left(\frac{\sqrt{3}q_y}{2}-\frac{q_x}{2}\right) \cr 
&+ \cos \left(\frac{q_x}{2}+\frac{\sqrt{3}q_y}{2}\right) + \cos(q_x)\biggr] \cr 
&+ \tau\sigma_2 \biggl[\sin \left(\frac{\sqrt{3}q_y}{2}-\frac{q_x}{2}\right) \cr 
&- \sin \left(\frac{q_x}{2}+\frac{\sqrt{3}q_y}{2}\right) + \sin(q_x) \biggr],
\end{align}
where $\tau$ denotes the hopping amplitude. 
The energy bands of graphene display Dirac cones in the points ${\mathbf{K}=(2\pi/3, 2\pi/3\sqrt{3})}$ and ${\mathbf{K}'=(2\pi/3, -2\pi/3\sqrt{3})}$ of the BZ. In the proximity of these points, the Hamiltonian takes the form ${\mathcal{H}(\mathbf{k})=v_F\mathbf{k}\cdot \boldsymbol{\sigma}}$, where ${\mathbf{k}=\mathbf{q}-\mathbf{K}}$ and ${v_f=3\tau/2}$.~Straightforward calculations show that $2{\boldsymbol{\mathcal{A}}(k_x,k_y)=(-k_y,k_x)/\sqrt{k_x^2+k_y^2}}$, which diverges in ${\mathbf{k}=0}$. However, the MCD is also affected by the $\sin^2(tE)$ factor, which approximates to ${\sin^2(tE)\sim 9\abs{\mathbf{k}}^2t^2/4}$ in the vicinity of the cone. The MCD is thus given by
\begin{equation}
    \boldsymbol{\mathcal{C}}(\mathbf{k},t)=\frac{9}{4}t^2\int_{BZ} \abs{G_{w,\mathbf{k}_0}(\mathbf{k})}^2(-k_y,k_x)\frac{dk_x,dk_y}{4\pi^2},
\end{equation}
which goes to zero for $\mathbf{q}\rightarrow \mathbf{K}$ (and similarly for $\mathbf{K}'$).
\begin{figure}[t]
    \centering
    \includegraphics[width=\columnwidth]{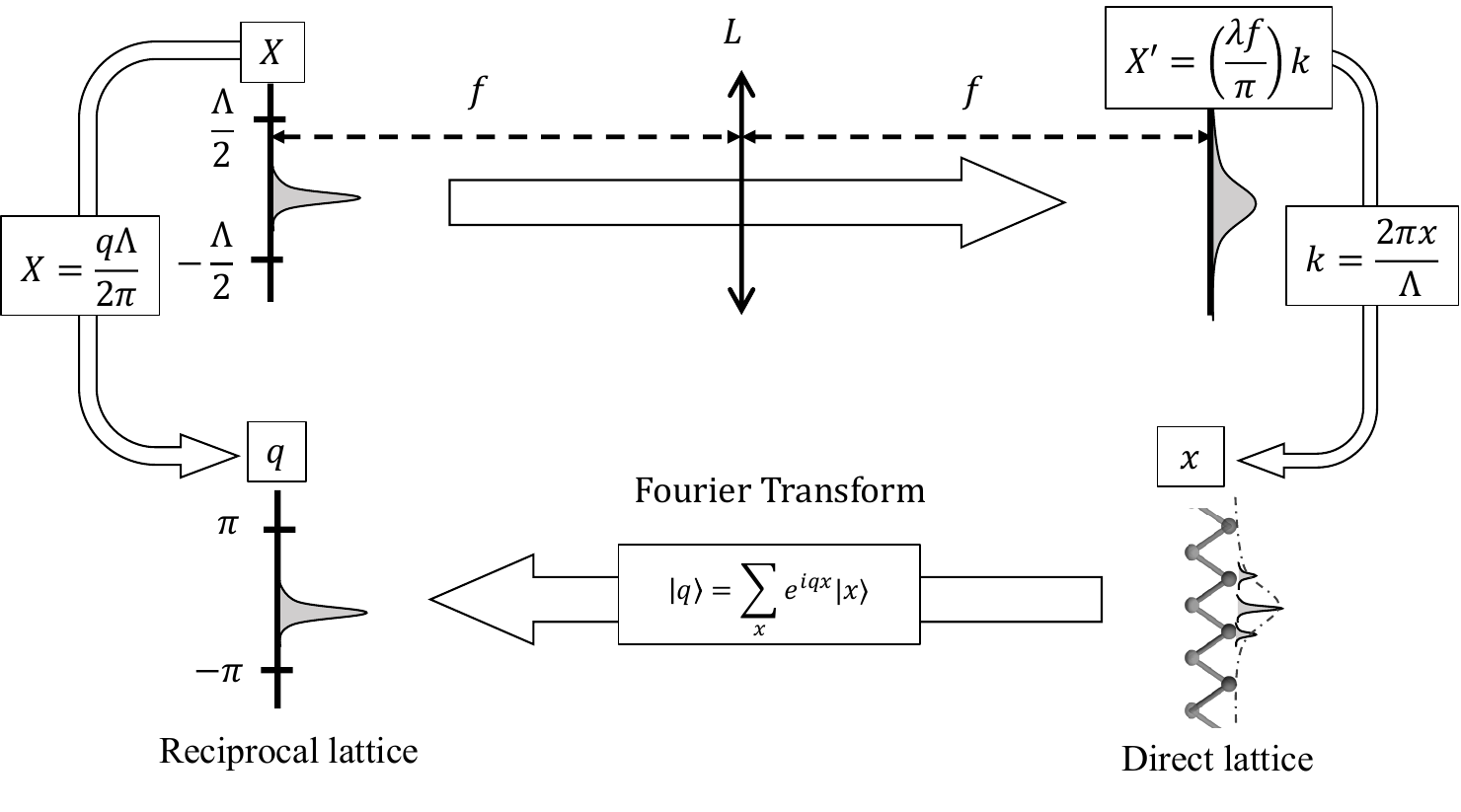}
    \caption{\textbf{Experimental encoding of the model parameters.} 
    The transverse coordinate $X$ in the plane of the metasurfaces is mapped into the reciprocal lattice coordinate $q$, and the characteristic distance $\Lambda$ corresponds to one Brillouin Zone. A narrow wavepacket in the reciprocal lattice thus corresponds to a beam crossing the $X$ plane with waist parameter ${w_0\ll\Lambda}$. The transverse coordinate $X'$ in the focal plane of the Fourier-transforming lens ($L$) is mapped in the lattice position $x$ of the QW via the relation $X'=2\lambda f x/\Lambda$. Note that $x$ and $q$ are considered adimensional quantities. }
    \label{fig:encoding}
\end{figure}

\subsection*{III. Details on the experimental setup}

The laser source is the output of a Titanium-Sapphire (Ti:Sa) laser (central wavelength 810 nm, pulse duration 150 fs, repetition rate 40 MHz), spatially cleaned through a single-mode fiber. The beam waist is ${w_0\simeq 2.5}$ mm to simulate localized initial states, and $w_0\simeq 0.32$ mm for wavepacket dynamics. 

\begin{figure}[t]
    \centering
    \includegraphics[width=\columnwidth]{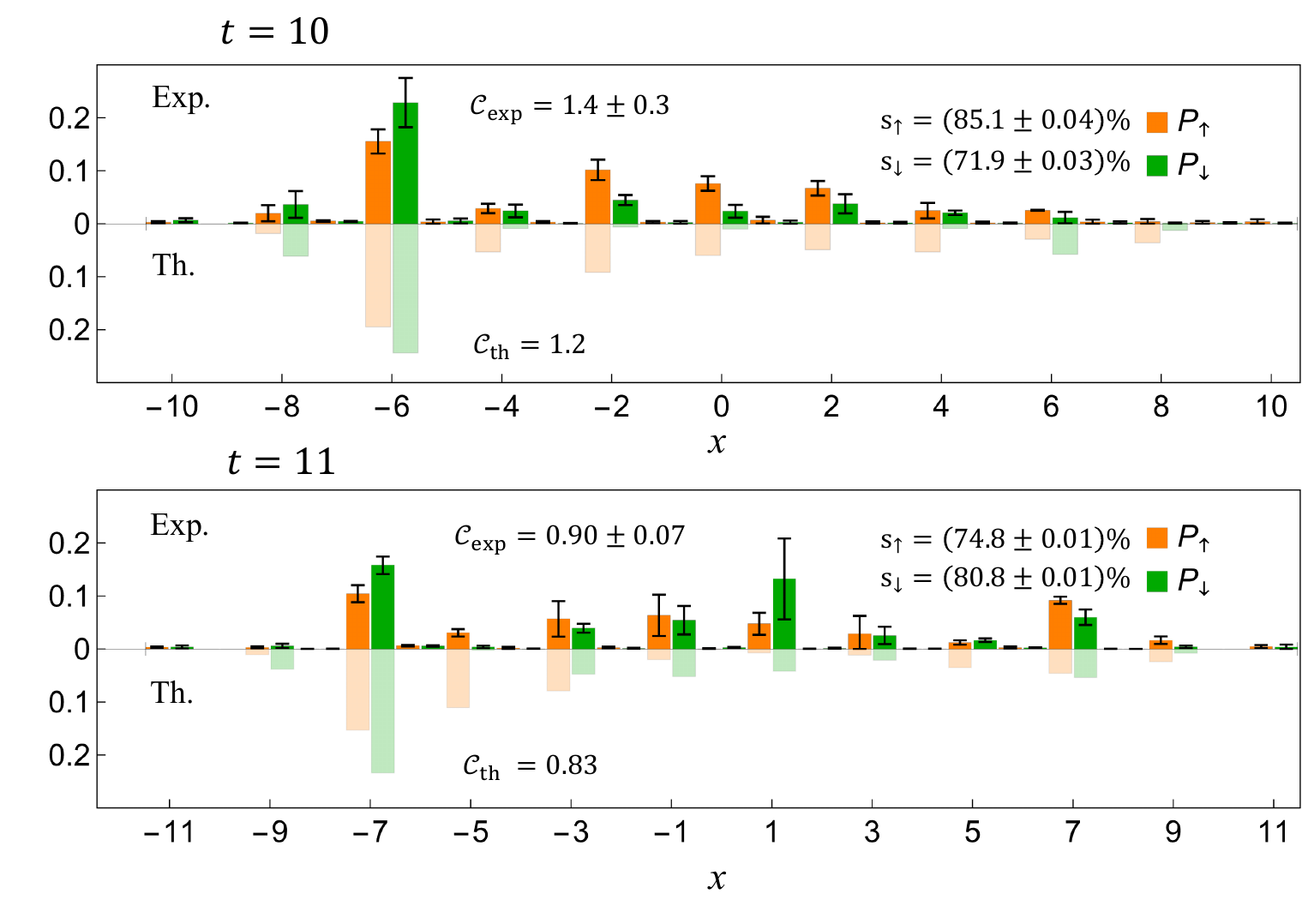}
    \caption{\textbf{Probability distributions for localized inputs.} Comparison between experimental and theoretical probability distributions after ${t=10}$ and ${t=11}$ steps for a $\ket{L}$-polarized localized initial state (${w_0>\Lambda}$), obtained after projecting on the chiral eigenstates $\ket{\uparrow}$ ($P_{\uparrow}$) and $\ket{\downarrow}$ ($P_{\downarrow}$). The experimentally measured MCD $\mathcal{C}_\text{{exp}}$, the theoretical prediction $\mathcal{C}_\text{{th}}$, and the similarity between the distributions are provided. Errors are reported as the mean standard error of four repeated measurements.}
    \label{fig:mcdloc}
\end{figure}

The liquid-crystal metasurfaces implementing the unitary evolution are fabricated with a photoalignment technique, based on orienting a dye solution (PAAD-22), spin-coated on ITO glasses, with linearly polarized light at 405~nm. The liquid crystal (6CHBT) is inserted in the sample via capillarity and locally aligns with the dye. Electrical contacts applied on the edges of the ITO surfaces allow tuning the optical retardation of the devices to the desired value. An alternate voltage with a sinusoidal wave at 10 kHz is used in the experiment.

\subsection*{IV. Relationship between experimental and simulated coordinate spaces}

The walker lattice space is encoded in the transverse-wavevector space of the light beam crossing the patterned waveplates. As illustrated in Fig.~\ref{fig:encoding}, this means that the transverse position $X$ (modulo $\Lambda$) in the plane of the liquid-crystal metasurfaces corresponds to the quasi-momentum $q$, while the far-field corresponds to the lattice space. Accordingly, a wavepacket corresponds to a beam having waist ${w_0<\Lambda}$ in the $X$ plane. The value $q_0$ can be controlled either by laterally shifting the wavepacket or, more practically, by translating the metasurfaces along the $X$ direction. The proper conversion factors from the setup to the model parameters are provided in Fig.~\ref{fig:encoding}.

\subsection*{V. Supplementary data}

We measure the QW distributions for localized input states after $t=10$ and $t=11$ time steps. The results are shown in Fig.~\ref{fig:mcdloc}. The agreement between the experimental observation and the theoretical prediction is quantified in terms of the similarity, ${s=(\sum_x\sqrt{P_\text{{exp}}(x) P_\text{{th}}(x)})^2}$, where $P_\text{{exp}}$ and $P_\text{{th}}$ are the normalized experimental and theoretical probability distributions, respectively.

\end{document}